\documentstyle[epsfig]{mn}

\def\Msun{\ifmmode{~{\rm M}_\odot}\else${\rm M}_\odot$~\fi}
\def\kms{\ifmmode{$~km\thinspace s$^{-1}}\else km\thinspace s$^{-1}$\fi}

\def\ee{\end{equation}}
\def\be{\begin{equation}}

\title{Luminosity limits on white dwarfs in a Galactic shroud}

\author[Janne Holopainen and Chris Flynn] 
{Janne Holopainen\thanks{e-mail: jaolho@utu.fi} and Chris Flynn\thanks{e-mail: cflynn@astro.utu.fi} \\
Tuorla Observatory, Piikki\"o, FIN-21500, Finland}

\date{}

\begin{document}
\maketitle

\begin{abstract}

We place observational constraints on a recently proposed Galactic population,
dubbed the {\it shroud} (Gyuk \& Gates 1999, Gates \& Gyuk 2001). The shroud would be a very
thick Galactic disk of low luminosity objects, most likely old white dwarfs,
proposed to explain the optical depth seen in microlensing surveys towards the
Magellanic clouds. The shroud is a simple alternative to the lenses being
distributed in a classical, near-spherical dark halo; the advantage of the
shroud is that it would compose only a fraction of a dark halo's total mass.

In this paper, we argue that stars of the Galactic shroud would be detectable
in the recent proper motion survey of Oppenheimer et al. (2001) if their
absolute luminosities were brighter than $M_{R_{59F}} = 19.4$ or approximately
$M_V = 18.6$. We adopt a range of simple models of the shroud's kinematics and
morphology, and the colours and luminosities of its white dwarfs; via
Monte-Carlo simulations, we predict the numbers expected in the Oppenheimer et
al. survey, which would be clearly separated from the numbers produced by white
dwarfs of the disk, thick disk and halo.

The number of white dwarf detections in the proper motion survey (98) is found
to be well explained by the disk, thick disk and halo. With {\it the most
conservative} kinematic and density parameters for the shroud, and an absolute
luminosity of the white dwarfs of $M_{R_{59F}} = 17.6$, we find that the proper
motion survey would detect over 100 WDs, just from the shroud.  For a
$M_{R_{59F}} = 19.4$ shroud, the survey would find $5 \pm 2$ peculiar objects,
whereas only two white dwarfs with such characteristics are found in the
original data. $M_{R_{59F}} = 19.4$ corresponds to $M_V = 18.6$ for WDs with
$(V - I) = -1.030$.

\end{abstract}


\section{Introduction}

The microlensing surveys carried out in recent years (e.g. EROS, MACHO and
OGLE) have reported on lensing population of dark objects seen towards the
Magellanic clouds. The favoured mass for these objects is approximately half a
solar mass, suggesting they are white dwarfs, since main sequence M stars of
this mass are bright enough to be detected directly in surveys. To date, no
directly detected counterpart for the dark population has been found. One
scenario is a population of ancient white dwarfs (WDs) which would comprise a
significant fraction of the Galactic dark halo (up to 20 per cent of its
mass). This suggestion has later become disfavoured because even quite dim WDs
would be directly detectable in the most recent proper motion surveys even if
they comprised `only' 2 per cent of the total mass of the dark halo (Reyl\'e,
Robin \& Cr\'ez\'e 2001; Flynn, Holopainen \& Holmberg 2003).

To explain the MACHO microlensing results (Alcock et al. 2000), and at the same
time to avoid some of the problems with a massive dark halo population, Gates
and Gyuk (hereafter, G\&G) (1999, 2001) proposed a new population in the Milky
Way, dubbed the Galactic `shroud'. It would be a WD population in the form of a
very thick disk with a scale height of 2.0 -- 3.0 kpc. It would produce the
same microlensing optical depth as a dark halo WD population without having to
be enormously massive.

We study here the implications of the shroud using the same techniques that we
used in a previous study (Flynn et al. 2003; hereafter, Paper I). In Paper I,
we constrained the luminosity and the number density of a dark halo WD
population with the same simulation that we use for the current study.  In the
previous study, we used two proper motion surveys independently for
constraining the halo population. After testing the simulation for Paper I, we
are now confident to use only the more recent one of those surveys (Oppenheimer
et al. 2001) for constraining the new population.

Oppenheimer et al. proper motion survey was originally designed to find dark
halo WDs. However, the follow up studies that have investigated the possibility
of dark halo WDs in this survey have found that the survey is also very
sensitive to the conventional thick disk WDs (e.g., Paper I; Reid, Sahu \&
Hawley 2001; Reyl\'e et al. 2001). Because the shroud has a similar velocity
structure to the thick disk, the survey is also very sensitive to the
shroud. Thus, the Oppenheimer et al. survey is optimal for our purposes.

In section 2, we briefly introduce how to find nearby WDs and separate them by
stellar population.  In section 3, we go through the parameters of the model in
detail, and in section 4, we describe the Oppenheimer et al. proper motion
survey and its findings. We illustrate the effect of proper motion to the
predicted number counts in section 5 and present our results in section
6. Finally, we conclude in section 7.

\section{White dwarf population separating techniques} \label{separating}

In our study, we are dealing with old (age $> 10$ Gyr) dim WDs, which have
cooled to surface temperatures of $3000 - 4000$ K and luminosities of less than
$1 \times 10^{-4}$ L$_\odot$. At present, detecting such white dwarfs in ground
based surveys is limited to the solar neighbourhood ($ < 100$ pc). This hampers
the detection of any density gradient with distance which could be used to
assign white dwarfs to various Galactic stellar populations. Consequently, the
population to which a local white dwarf belongs can only be determined from its
kinematics, since the different populations have different rotation velocities
and velocity dispersions. The exception is for younger white dwarfs of the
disk; being younger these WDs are brighter, and their luminosities assist in
assigning a population type (for a detailed review on identifying WDs, see
Hansen \& Liebert 2003).

Field white dwarfs can be extracted from catalogs via their proper motions,
magnitudes and colours, or the combination of apparent magnitude and proper
motion termed the `reduced proper motion', $H_X$. It is defined as

\be
H_X = X + 5 log(\mu) + 5 = M_X + 5 log(V_{\mathrm tan}) - 3.38,
\ee

where $X$ is the apparent magnitude, $\mu$ is the proper motion in
arcseconds per year, $M_X$ is the absolute magnitude and $V_{\mathrm
tan}$ is the tangential space motion in km s$^{-1}$.

Extremely high values of the reduced proper motion (for example, $H_V \sim 24$)
indicate stars which are intrinsically dim and fast moving; both criteria of
interest when searching for dim stars which would make up either a dark halo or
dark shroud around the Galaxy.

At present, the state-of-the-art in finding intrinsically dim WDs is via proper
motion surveys, selection of the high reduced proper motion objects, followed
by detailed spectroscopy of all the candidates (for a list of the latest
studies, see Hansen \& Liebert 2003). The best survey to date is that of
Oppenheimer et al. (2001), which has been augmented later by Salim et
al. (2003).  The two groups have identified some 98 dim white dwarfs in a 4000
square degree survey at the SGP, reaching to $V \sim 20$.

\section{The model parameters of the shroud}

\subsection{A new very thick disk}

The shroud is a {\it very thick} Galactic disk and has a scale height two to
three times higher than the conventional thick disk. The main motive for
proposing the existence of the shroud is a need to explain a microlensing
optical depth $\tau \sim 10^{-7}$ towards the LMC (Alcock et
al. 2000). This value can be explained with a massive dark halo of
WDs. However, the progenitor stars of this population would require a highly
peaked IMF and should pollute the Galaxy with metals (e.g., Brook et al. 2003).

Explaining the optical depth with an extended thick disk instead of a dark halo
offers an alternative solution. This is because the required total mass of the
population can be brought down from the dark halo's $\sim 10^{12}$ M$_{\odot}$
to the shroud's $\sim 6 \times 10^{10}$ M$_{\odot}$. This alleviates the
pollution problem, although certainly not entirely.

G\&G describe the shroud in detail in Gyuk \& Gates 1999 (GG99) and Gates \&
Gyuk 2001 (GG01). The first paper constrains the model parameters, and in the
second paper, G\&G propose that the shroud consists of ancient white dwarfs. We
examine the white dwarf scenario, which is based on the general models
presented in GG99 (this paper is also the source for more detailed description
of the models for the interested reader). We adopt the parameters and
constraints given in GG99, and then, we explore how these constraints and the
capabilities of the proper motion survey together limit the number of
detectable shroud WDs.

\subsection{Density structure}

In GG99, G\&G give two radial density models for their extended thick disk: 

\be \Sigma(r) = \Sigma_0 {\rm exp}[(r_0 - r)/r_d] \ee

and 

\be \Sigma(r) = \Sigma_0 \frac{r_0 + a}{r + a}, \ee

where the first is a model similar to the standard thick disk and the second
resembles a flattened halo. Here, $\Sigma$ is the surface density, $\Sigma_0$
is the surface density at the Sun's Galactocentric radius, $r$ is the
Galactocentric radius, $a$ is a core radius, $r_d$ is the scale length of the
extended thick disk and $r_0$ is the Galactocentric radius of the Sun.
Thus, the models are parametrized by surface density, $\Sigma_0$,
which is typically in the range

\be 
60 \thinspace {\rm M}_{\odot} {\rm pc}^{-2} \leq \Sigma_0 \leq 115
\thinspace {\rm M}_{\odot} {\rm pc}^{-2}.  
\ee

G\&G adopt two density distributions for the extended thick disk as a
combination of the radial direction (exponential or $(r + a)^{-1}$) and the
vertical direction (sech$^2$) distributions:

\be
\rho(r,z) = \frac{\Sigma_0}{2h_z} {\rm exp}[(r_0 - r)/r_d] {\rm  sech}^2(z/h_z)
\ee

and 

\be
\rho(r,z) = \frac{\Sigma_0}{2h_z} \frac{r_0 + a}{r + a} {\rm  sech}^2(z/h_z)
\ee

We are interested only in constraining the local density at the Sun $\rho_0 =
\rho(r_0,0)$, i.e. $\rho_0 = \frac{\Sigma_0}{2h_z}$. We consider scale heights
in the range
\be
2.0 \thinspace {\rm kpc} \leq h_z \leq 3.0 \thinspace {\rm kpc}, 
\ee
following the constraints from GG99.

The modeled shrouds are parametrized by surface density, $\Sigma_0$ and scale
height, $h_z$. These two parameters form a parameter space which is constrained
by observational values of i) optical depth, $\tau \geq 10^{-7}$; ii) rotation
velocity of the shroud, $v_c \leq 180$ \kms and iii) total vertical column
density of the disk within 1.0 kpc, $\Sigma_{tot, 1.0} \leq 90$ M$_{\odot}$
pc$^{-2}$ (for references, see GG99). By combining the ranges of $\Sigma_0$ and
$h_z$, the local mass density range becomes

\be 
0.010 \thinspace {\rm M}_{\odot} {\rm pc}^{-3} \leq \rho_0 \leq 0.029
\thinspace {\rm M}_{\odot} {\rm pc}^{-3}.  
\ee

We adopt a WD mass of $m_{WD} = 0.6$ \Msun, leading finally to the space
density of nearby WDs, $\rho_n$, in the range

\be
0.0167 \thinspace {\rm stars} \thinspace {\rm pc}^{-3} \leq \rho_n \leq 0.0483
  \thinspace {\rm stars} \thinspace {\rm pc}^{-3}.
\ee

\subsection{Velocity structure}

G\&G do not restrict the rotation velocity of their model too much. They allow
a range $v_c = 130 - 180$ \kms. We use these values as our lower and upper
limit. We use a solar rotation velocity of $v_{\odot} =$ 220 km s$^{-1}$, and
the $v_c$ values convert to an asymmetric drift values in the range $-90$ \kms
to $-40$ \kms. The vertical velocity dispersion is that of an isothermal disk:

\be
\sigma_W = \sqrt{2 \pi G \rho_0 h_z^2} = \sqrt{\pi G \Sigma_0 h_z}
\ee 

Furthermore, G\&G use a typical dependence in a disk population for the other
dispersions (e.g. Binney \& Tremaine 1987): $\sigma_r \approx \sqrt{2}
\sigma_W$ and $\sigma_{\phi} \approx \sigma_W$.

From the limits on $\Sigma_0$ and $h_z$ above, we derive limits on the velocity
dispersions of 57 \kms $\leq \sigma_U \leq$ 99 \kms, 40 \kms $\leq \sigma_V
\leq$ 70 \kms and 40 \kms $\leq \sigma_W \leq$ 70 \kms.

\subsection{Luminosities of WDs} \label{luminosity structure}

For cool, old, Hydrogen atmosphere white dwarfs (ages 10 to 14 Gyr), the
expected luminosity range is roughly $16 \leq M_V \leq 18$ (Richer 2000, Hansen
2001). Models of WDs with ages of 13 -- 14 Gyr which would have cooling curves
down to $M_V = 19$ are not presently favoured, although they can be
constructed. However, $M_I$ values are not so well restricted because the
cooling curves allow a large range of $M_V, (V - I)$ -dependencies.

G\&G constrain the luminosity of the shroud in the $I$-band, based on counts of
faint sources in the Hubble Deep Field, whereas we use the $R_{59F}$-band of
the Oppenheimer et al. survey. We do not constrain the shroud based on the
$I$-range of G\&G but on the cooling curves and evolution models of WDs.

Realistically, any dim ancient white dwarf population must have a present day
luminosity function (LF) which is a reflection of the initial mass function in
the WD progenitors. Simulating the star formation and evolution process via
isochrones and cooling curves is no simple matter. We do not attempt to test
here plausible LFs other than a simple delta function; i.e. all the WDs have
the same luminosity.  It turns out that this is the most conservative option
one can adopt, as will be discussed in Section \ref{rpm window}.

\section{The proper motion survey}

The Oppenheimer et al. (2001) (hereafter, OHDHS) proper motion survey is the
most effective WD proper motion survey to date. The survey is complete to a
detection limit of $R_{59F} < 19.7$, and the proper motion detection window is
0.33 arcsec yr$^{-1} \leq \mu \leq 3.0$ arcsec yr$^{-1}$. The survey is most
likely partially incomplete at the upper proper motion boundary. We make
conservative assumptions in order to take this into account in Section
\ref{res3}. The survey covers 10 per cent of the sky, about 4000 square degrees
at the SGP.

OHDHS found 38 high velocity white dwarfs among 98 WDs in total. The high
velocity WDs were provisionally assigned to the dark halo by OHDHS; their data
implied that some 2 per cent of the dark halo was in this (baryonic) form. In
later studies (e.g., Paper I; Hansen \& Liebert 2003; Reid, Sahu \& Hawley
2001), most of the high velocity WDs have been assigned to the traditional
thick disk and stellar halo. There are only 2 WDs in the survey that have
reduced proper motion characteristics ($H_{R_{59F}} > 24$) which are not
typical for any of the known populations.

In a follow up study, Salim et al. (2003) studied the 38 high velocity WDs in
more detail using photometry and spectroscopy. They calibrated the photometric
$R_{59F}$--band used by OHDHS to standard $V$. In Paper I, we used a similar
colour transformation which was based on an empirical calibration of the
$R_{59F}$--band to $R$ using M dwarfs. Salim et al.'s transformation is quite
close to the one we adopted in Paper I.

We use the $R_{59F}$ calibration by Salim et al. at the end of this paper
because it depends strongly on the $V - I$ colours of the WDs. Because $V - I$
colours for a WD population vary according to the WD type, age, mass and
evolution model, the conversion from $R_{59F}$ to standard magnitudes results
to a wide range of $M_V$ values for the population. Thus, we construct the
shroud with absolute magnitude values in $M_{R_{59F}}$ and also simulate the
OHDHS survey in this original band. We examine the transformation to
$M_V$-magnitudes of $M_{R_{59F}}$ for a full range of plausible models in
Section \ref{res_v}.

\begin{figure}
\begin{center}
\epsfig{file = 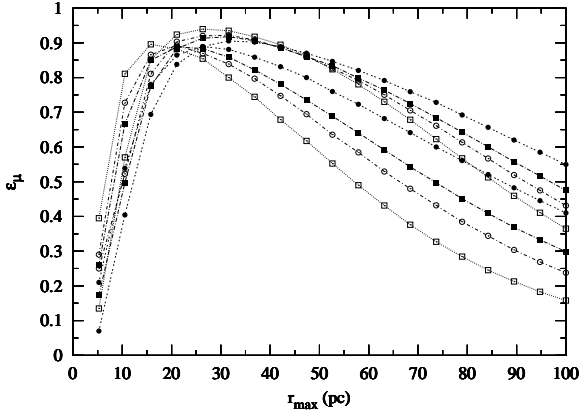, angle = -90, width = 80mm}
\caption[]{The $\mu$--completeness of the survey for the shroud up to 100
pc. The differences between the number counts for the models are produced by
differences in the parameters: $r_{max}$, $\sigma_W$ and $v_c$. See Table 1 for
model identification.}
\label{e_d_100pc}
\end{center}
\end{figure}

\begin{figure}
\begin{center}
\epsfig{file = 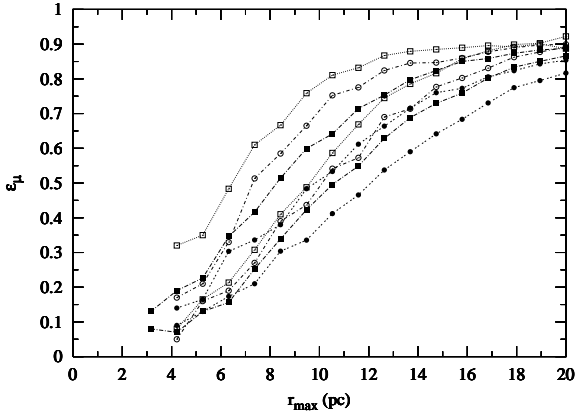, angle = -90, width = 80mm}
\caption[]{As for Figure \ref{e_d_100pc}, but 0 pc $ < r_{max} < $ 20 pc. The
$\mu$--completeness of the survey is close to 90 per cent for the shroud with a
detection distance of 20 pc. See Table 1 for model identification.}
\label{e_d_20pc}
\end{center}
\end{figure}

\section{Counting the shroud white dwarfs}

\subsection{Proper motion window}

The number of stars detectable within a proper motion window of a survey is:

\be
N_\mu = \rho_n V \varepsilon_\mu,  
\ee

where $\rho_n$ is the number density of the population, $V$ the survey volume
and $\varepsilon_\mu$ the fraction of stars which are inside the proper motion
window. For a test population of a given luminosity, $M_{R_{59F}}$, the volume is
determined by the maximum detection distance for that population, $r_{max}$.

For the shroud, $\varepsilon_\mu$ is a function of the maximum detection
distance ($r_{max}$), $W$-velocity dispersion ($\sigma_W$) and rotation
velocity ($v_c$). Within the parameter ranges, $\varepsilon_\mu$ depends
strongly on all three parameters.

We can interpret $\varepsilon_\mu$ as the $\mu$-completeness of a survey -- it
gives the probability that an object from a given population has a proper
motion inside the proper motion window of the survey. In our case,
$\varepsilon_\mu$ is the OHDHS survey $\mu$-completeness for a given shroud. It
is a convenient quantity because it is independent of the number density of the
population.

The solid angle of the OHDHS survey is 10 per cent of the sky and results to a
volume of:

\be
V = \frac{\Omega}{3} r_{max}^3 = 0.419 \thinspace r_{max}^3.
\ee
Now the equation for $N_\mu$ can be written:
\be \label{N_mu} 
N_\mu = 0.419 \thinspace r_{max}^3 \thinspace \rho_n(h_z, \Sigma_0) \thinspace
\varepsilon_\mu(r_{max}, v_c, \sigma_W) 
\ee

\subsection{Reduced proper motion window} \label{rpm window}

As mentioned in Section \ref{separating}, a high $H$ value indicates that the
object is faint and has a large tangential space velocity. If we want to
separate a peculiar population (like the shroud) from the known populations,
one way is to search for unusually high $H$ values. The OHDHS survey data shows
that a suitable limit which will exclude the known populations effectively is
$H_{R_{59F}} > 24$. For example, for a $M_{R_{59F}} = 16$ object, $V_{tan}$
must be larger than 188 \kms\ to satisfy this limit.

The number of WDs which have extremely high reduced proper motion values can be
expressed

\be 
N_{\mu H} = 0.419 \thinspace r_{max}^3 \thinspace \rho_n \thinspace
\varepsilon_{\mu H}, 
\ee

which is analogous to Equation \ref{N_mu}. $\varepsilon_{\mu H}$ acts as
$\varepsilon_\mu$, only now it is the fraction of WDs which have 0.33 arcsec
yr$^{-1} \leq \mu \leq 3.0$ arcsec yr$^{-1}$ {\it and} $H_{R_{59F}} > 24$.

As mentioned in Section \ref{luminosity structure}, we adopt a delta function
for the LF for all our models. This procedure increases $N_{\mu H}$ while it
minimizes $N_{\mu}$. The fact that $N_{\mu H}$ is increased might be a problem
because we use $N_{\mu H}$ to constrain shrouds which are extremely faint. To
resolve the dependency between $N_\mu$ and $N_{\mu H}$ as a function of the
shape of the LF, we ran simulations with 2 magnitude wide top-hat LFs and
compared the results to models with delta function LFs; i.e. the top-hat LFs
contained stars distributed equally up to two magnitudes brighter than those in
the delta function LF. As expected, wide LFs always produce much higher
$N_{\mu}$ values than a delta function LF, and we were able to constrain even
the faintest shrouds only by $N_{\mu}$. Thus, adopting a faint delta LF is the
most conservative method for all our models, because it produces the smallest
$N_{\mu}$ values for a given model.

\begin{figure}
\begin{center}
\epsfig{file=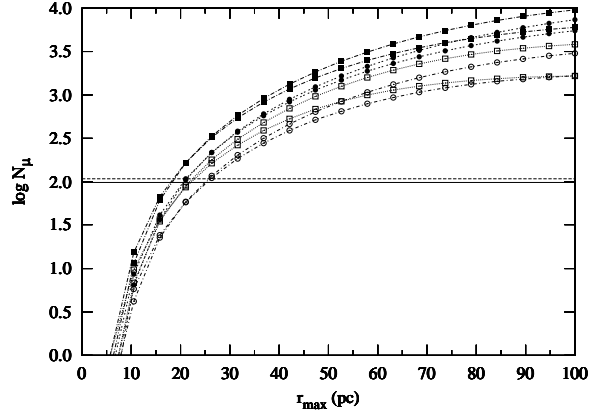, angle=-90, width=80mm}
\caption[]{The number of shroud WDs (in logarithmic scale) which are inside the
detection volume and have proper motion values 0.33 arcsec yr$^{-1} \leq \mu \leq
3.0$ arcsec yr$^{-1}$. See Table 1 for model identification.}
\label{n_d_100pc}
\end{center}
\end{figure}

\begin{figure}
\begin{center}
\epsfig{file=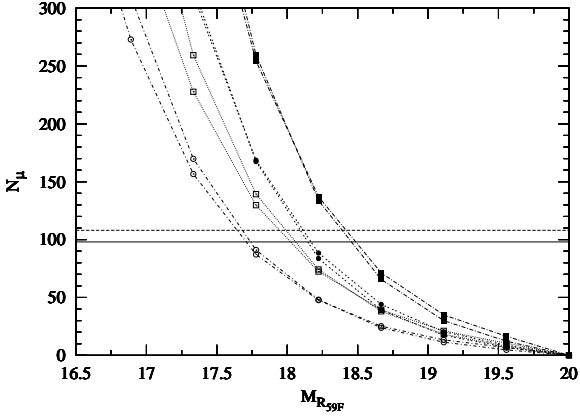, angle=-90, width=80mm}
\caption[]{Another way to look at the results in Figure \ref{n_d_100pc} with
$r_{max}$ converted to the absolute magnitude of the shroud with the limiting
magnitude $R_{59F}^{lim} = 19.7$. See Table 1 for model identification.}
\label{n_MR1620}
\end{center}
\end{figure}

\begin{figure}
\begin{center}
\epsfig{file = 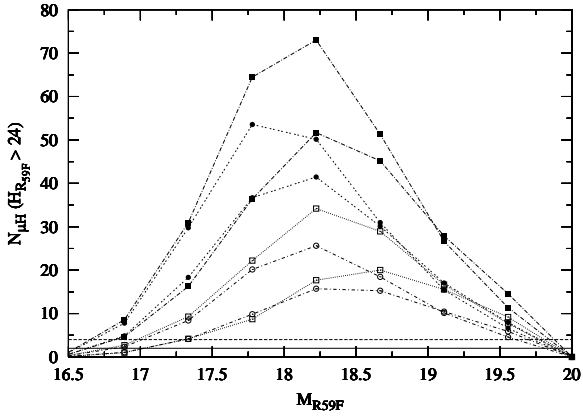, angle = -90, width = 80mm}
\caption[]{The number of shroud WDs which have a reduced proper motion value
$H_{R_{59F}} > 24$. OHDHS detected 2 such WDs. See Table 1 for model
identification.}
\label{nHR24_MR1620}
\end{center}
\end{figure}

\section{Results} \label{res}  

\subsection{Limiting the shroud by $N_\mu$} \label{res1}

We limit the number of detectable objects from the shroud in the OHDHS survey
to $N_\mu^{max} = 108$. This is the 1-$\sigma$ upper limit on the number of WDs
which the OHDHS survey actually detected (98).  The limits are shown as
horizontal lines in Figures \ref{n_d_100pc} and \ref{n_MR1620}.

In our highly conservative scenario, we assume that all the 98 WDs detected by
the OHDHS survey are from the shroud and none are from the other Galactic
populations. We do not attempt to subtract the disk component from these WDs,
although it is highly probable that $\sim$ 60 of the 98 WDs are from the thin
disk based on their velocity components. Fortunately, $N_{\mu}^{max} = 108$
still sets some interesting limits on the luminosity of shroud WDs.

Figures \ref{n_d_100pc} and \ref{n_MR1620} show the expected number of WDs in
OHDHS for various shroud models and adopted WD luminosities.  Changing the
rotation velocity of the shroud affects the number counts considerably less
than the changes in $h_z$ and $\Sigma_0$: Each line style represents two models
which are separated only by $v_c$-differences ($v_{c1} = 130$ km s$^{-1}$,
$v_{c2} = 180$ km s$^{-1}$). Figure \ref{n_MR1620} shows clearly that the
number counts are nearly equal in both cases.

Not surprisingly, the model which produces the lowest $N_\mu$ values has the
lowest number density, $\rho_n = 0.0167$ stars pc$^{-3}$. The model can be seen
in Figure \ref{n_MR1620} where it produces 108 WDs when its luminosity is
$M_{R_{59F}} = 17.6$ or $17.7$, depending on its rotation velocity. Thus, we
consider models with $M_{R_{59F}} < 17.6$ to be so luminous that they can be
ruled out directly by proper motion based number counts in the OHDHS survey.
    
\begin{table}
\begin{center}
\caption{Symbols and parameters of the models presented in the figures.}
\begin{tabular}{ccccc}
\hline
\hline
Symbol & $h_z$ & $\Sigma_0$ & $\sigma_W$ & $v_c$ \\
& kpc & \Msun pc$^{-2}$ & km s$^{-1}$ & km s$^{-1}$\\
\hline 
$\sq$ & 2.0 & 60.0 & 40.3 & 130.0, 180.0 \\
$\circ$ & 3.0 & 60.0 & 49.3 & 130.0, 180.0 \\
\rule{1.3ex}{1.3ex} & 2.0 & 115.0 & 55.7 & 130.0, 180.0 \\
$\bullet$ & 3.0 & 115.0 & 68.3 & 130.0, 180.0 \\
\hline
\end{tabular}
\end{center}
\end{table}
        
\subsection{Limiting the shroud by $N_{\mu H}$} \label{res2}

The fact that the OHDHS survey was able to find only 2 WDs with $H_{R_{59F}} >
24$ allows us to constrain the shroud to even fainter models than with the
proper motion window.

We consider a model to be ruled out when it produces $N_{\mu H} > 4$ WDs. The
original count of the OHDHS survey and the Poissonian 1-$\sigma$ error limit
are shown as horizontal lines in Figure \ref{nHR24_MR1620}.

Again, it turns out that the most conservative model at the low luminosity end
is the one with the lowest number density. As in Section \ref{res1}, the
luminosity of the shroud is limited to nearly equal $M_{R_{59F}}$ values
regardless of the rotation velocity of the limiting model. The model can be
seen in Figure \ref{nHR24_MR1620} where it produces 4 WDs when the WD
luminosity is at either $M_{R_{59F}} = 17.0$ or 19.6. Thus, we consider models
with $17.0 < M_{R_{59F}} < 19.6$ to be such that they can be ruled out directly
by reduced proper motion based number counts in the OHDHS survey. This,
together with the results presented in Section \ref{res1}, rules out all the
models which have $M_{R_{59F}} < 19.6$.

\subsection{Proper motion incompleteness effect} \label{res3}
 
In Sections \ref{res1} and \ref{res2}, we have assumed that the OHDHS survey is
100 per cent complete to the upper limit of the proper motion window,
$\mu_{max} = 3.0$ arcsec year$^{-1}$.  OHDHS conduct a completeness test to
resolve their brightness detection limit, but the completeness of their proper
motion window is not fully resolved.

OHDHS estimate that stars with proper motions more than 3 arcsec year$^{-1}$
have only 10 per cent chance to be found in their survey. Due to this
uncertainty, we conducted additional simulations in which we assumed that the
survey would be complete to only $\mu_{max} = 2.0$ arcsec year$^{-1}.$ It turns out that
this does not affect the results of Section \ref{res1} because most of the WDs have proper motions
less than 2.0 arcsec year$^{-1}.$ For Section \ref{res2} results, adopting the lower proper
motion limit changes the most conservative brightness limit
from $M_{R_{59F}} = 19.6$ to $M_{R_{59F}} = 19.2$ for the $v_c = 130$ km s$^{-1}$ shroud.
In the case of $v_c =180$ km s$^{-1}$, the limit is changed by only 0.2 magnitudes, to $M_{R_{59F}}
= 19.4$.

To further understand the effect of the proper motion window, we ran
simulations with a changing completeness level. We adopted a linear drop in the
completeness level from 100 to 0 per cent for the range $1.5 {\rm \thinspace arcsec \thinspace year}^{-1} <
\mu < 3.1 {\rm \thinspace arcsec \thinspace year}^{-1}.$ This resulted in smaller effects than adopting a
crude cut of $\mu_{max} = 2.0$ arcsec year$^{-1}$ and changed the brightness
limit at most from $M_{R_{59F}} = 19.6$ to $M_{R_{59F}} = 19.4$.

In the most conservative model, WDs with $H_{R_{59F}} >24$ have a mean proper
motion $\bar{\mu} \sim 1.3$ arcsec year$^{-1}$. Thus, adopting $\mu_{max} = 2.0$
arcsec year$^{-1}$ instead of $\mu_{max} = 3.0$ arcsec year$^{-1}$ for an upper
limit does not have a large effect on the results presented in Sections \ref{res1}
and \ref{res2}. We consider the limit $M_{R_{59F}} = 19.4$ derived above to be
very conservative within the given parameters. 
 
\subsection{Limiting the shroud luminosity in $V$-band} \label{res_v}

Finally, we want to transform the result presented in Section \ref{res3} to
standard magnitudes.  This is needed to link our models to general evolutionary
models of WDs and evaluate how probable the shroud scenario is. We use the
Salim et al. (2003) transformation between $R_{59F}$ and $V$. This is based on
photometry of 17 WDs which have been observed in the OHDHS survey.  The
transformation can be written

\be
\label{salim} V - R_{59F} = 0.66 (V - I) - 0.13, 
\ee 

and has a scatter in $R_{59F}$ of $\sigma_{R_{59F}} = 0.09$. As mentioned in
Section \ref{luminosity structure}, this transformation is close to the
transformation we used in Paper I (given here as a linear fit):

\be
V - R_{59F} = 0.52 (V - I) - 0.02.
\ee

The Salim et al. transformation is superior because it is based on actual
WDs, rather than our earlier adopted transformation which was based on M 
dwarfs (Bessell 1986); no such WD data were available to us at the time.

We want to estimate a $M_V$-limit which would correspond to $M_{R_{59F}} =
19.4$ for the shroud WDs. From Equation \ref{salim}, we see that this requires
some $V - I$ range evaluation.

Salim et al. have measured or derived $V - I$ values for a subsample of the
OHDHS WDs. The lowest value they can find is for LHS 1402, which has $V - I =
-0.37$. We can rule out shrouds with $M_V < 19.0$ by adopting this value.  It
must be noted that LHS 1402 is not a typical WD in this sample (although it
might be a fair representative of a shroud WD); adopting redder colours, from
the other WDs, would result in even fainter limits.

We can also use the cooling curves of Richter et al. (2000) which for the
faintest WDs predict $V - I$ values down to $-1.030$. This is for a 0.6 \Msun
mass WD with an age of 15 Gyrs and a luminosity of $M_V = 18.1$. Adopting $V
- I = -1.030$ rules out shrouds with $M_V < 18.6$.

G\&G limit the shroud in $I$-band to $M_I \sim 16 - 17$. The above $V - I$ and
$M_V$ values can also be used to limit the shroud in $I$: The first estimate,
$V - I = -0.37$, rules out shrouds with $M_I < 19.6$. The second value, $V - I
= -1.030$, sets similar limits, shrouds with $M_I < 19.4$ are ruled out. The
lowest limit in $I$ can be found by using $V - I = 2.0$. This rules out shrouds
with $M_V < 20.6$ and $M_I < 18.6$.

\section{Conclusions}

Strong limits have been placed on the luminosities of white dwarfs which could
make up the putative `shroud' of the Galaxy, proposed as a solution to the
optical depth measurements seen in the microlensing surveys towards the
Magellanic clouds. We use the Oppenheimer et al. (2001) proper motion survey of
4000 square degrees to $R_{59F} = 19.7$, containing 98 spectroscopically
confirmed WDs. A range of shroud models are investigated, and the number of WDs
with high reduced proper motions compared to the Oppenheimer et al. data. Most
of the models produce significantly more WDs than are actually observed; in
particular, models in which we probe the very highest reduced proper motion
source (indicating very low luminosity and high space velocities, as expected
for the shroud component) allow us to limit the luminosity of the WDs in the
shroud to $M_{R_{59F}} = 19.4$ (the survey pass band), which corresponds to
$M_V = 18.6$ or $M_I = 19.6$. If the Galaxy possesses a shroud of WDs which
produces the microlensing signal, these WDs must be fainter than the above
limits. Very few models of Hydrogen atmosphere WDs cool to such faint levels
within the age of the Universe.

\section*{Acknowledgments}

This research was supported by the Academy of Finland through its funding of
the ANTARES program for space research. CF thanks the Centre for Astrophysics
and Supercomputing, Swinburne University of Technology for support during a
visit where part of the research was carried out. JH thanks the V\"ais\"al\"a
Foundation and Swinburne for financial support and travel assistance.

\end{document}